\documentclass[11pt,twocolumn,twoside]{article}
\usepackage{iccr2024}

\usepackage[export]{adjustbox}

\begin{document}

\title{IQMDose3D: a software tool for reconstructing the dose in patient using patient planning CT images and the signals measured by IQM detector} 

\author[1,2,3]{Aitang Xing}
\author[1,2,3]{Gary Goozee}
\author[1,2,3]{Alison Gray}
\author[1,2,3]{Vaughan Moutrie}
\author[1,2,3]{Sankar Arumugam}
\author[1,2,3]{Shrikant Deshpande}
\author[1,2,3]{Anthony Espinoza}
\author[1,2,3]{Vasilis Kondilis}
\author[1,2,3]{Marjorie McDonald}
\author[1,2,3]{Philip Vial}

\affil[1]{Liverpool \& Macarthur Cancer Therapy Centre, Liverpool, NSW 1871, Australia}

\affil[2]{Ingham Institute for Applied Medical Research, Liverpool, NSW 1871, Australia}

\affil[3]{South West Sydney Clinical Campuses, University of New South Wales, Sydney, Australia}

\maketitle
\thispagestyle{fancy}


\begin{customabstract}
The integral quality monitor (IQM) system compares the signal measured with a large volume chamber mounted to the linear accelerator’s head to the signal calculated using the patient DICOM RT plan for patient-specific quality assurance (PSQA). A method was developed to reconstruct the dose in patients using the signal measured by IQM chamber and patient planning CT images. A software tool named IQMDose3D was implemented to automate this procedure and integrated into the IQM-based PSQA workflow. IQMDose3D enables the physicists to evaluate PSQA by focusing on the clinical perspective by comparing the delivered plan to the approved clinical plan in terms of the clinical goals, dose-volume histogram (DVH)  in addition to the three-dimensional (3D) gamma map and gamma pass rate. 
\end{customabstract}


\section{Introduction}
The Integral Quality Monitor (IQM) system is a commercial product from iRT Systems GmbH (Koblenz, Germany) designed for patient-specific quality assurance(PSQA). The unique feature of the system is the IQM detector which can be mounted to linear accelerator (Linac) \cite{Ghafarian2021}. The IQM detector is a large-volume wedged-shaped ionization chamber. The wedged-shaped design makes it very sensitive to the position of the multileaf collimator(MLC) \cite{Alharthi2021}. The system’s other advantages include no need to create a QA plan and set up a phantom before measurement and automated generation of a PDF report after measurement. IQM was commissioned and has been used for routine PSQA for patients planned in the RayStation treatment planning system (TPS) since 2022 in our centre \cite{Gray2023,Michel2021}. In this study, a method was proposed to reconstruct the dose in patients using the signal measured by the IQM detector. A software tool was developed to automate this procedure and integrated into the current IQM-based PSQA workflow.  

\section{Materials and Methods}

The IQM detector can measure the accumulated and segment-based signal for each segment of the VMAT beams. The measured signal is compared to the signal calculated using the patient’s plan by an IQM model established during the commissioning. The percentage differences between the calculated and measured segment-by-segment and accumulated signals were calculated and used to determine if the PSQA failed or passed using criteria established during the commissioning \cite{Xing2023,IRTManual}. The calculated signal can be mathematically expressed as \cite{IRTAlgorithm}:
\begin{equation}
   S_{calc}=MU_{seg}\times K \times AOF\left ( x,y \right ) \times F\left ( x,y \right )_{calc}
   \label{eq:s_calc}
\end{equation}
In the equation (\ref{eq:s_calc}),  \(MU_{seg}\) is the monitor unit (MU) for a  segment, and K is constant. \((x,y)\) is the position of the elementary beamlet for this segment.  \(AOF(x,y)\) is the area-integrated output factor and is calculated via a look-up table established during the commissioning. \(AOF(x,y)\)  is a constant for a segment with a fixed shape and location. \(F(x,y)_{calc} \)is the fluence map calculated using the photon source model implemented in the IQM system, MLC and jaw transmission and integrated over the shape of the segment. For a segment with a fixed shape and location, it is kept unchanged.  

The measured signal can be expressed as:
\begin{equation}
   S_{meas}=MU_{meas}\times K \times AOF\left ( x,y \right ) \times F\left ( x,y \right )_{calc}
\end{equation}
Where \(MU_{meas }\)  is the delivered segment MU. Although theoretically \(AOF(x,y) \)and \(F(x,y)_{calc} \) are constant for a segment with a fixed shape and location, they may be slightly changed due to the actual MLC or Jaw positions or machine output fluctuation during the delivery. Therefore  \(MU_{meas } \) is an effective MU taking into account these changes. For each segment, the effective \(MU_{meas}\) can be calculated as:

\begin{equation}
    MU_{meas}=MU_{cal}\times \frac{S_{meas}}{S_{calc}}
\end{equation}

Once the measured MUs are calculated for each segment, they can be imported back to Raystation to reconstruct the dose in the patient and compared to the original plan. A software tool named IQMDose3D was developed to automate this procedure. 

IQMDose3D was developed using Python and the scripting API of Raystation. The software tool was divided into three modules: (1) IQM signal module. This module is responsible for processing IQM-generated Excel files containing the calculated and measured signal. It not only calculates the \(MU_{meas}\), but also queries other patient or plan information. (2) Raystation module. The module uses the Raystation scripting to perform the following tasks sequentially: firstly, a plan named IQM was created by copying the clinical plan and applying the measured segment MUs; secondly, a 3D gamma map was calculated using the original clinical plan and IQM plan. The third plan named Gamma was created by copying the original clinical plan, but the dose was replaced with a 3D gamma map. (3) Report module. After the user manually reviews the IQM plan and clinical plan as well as the 3D gamma map, the module is used to generate a PDF report for PSQA. 

\begin{figure}
  \centering
  \includegraphics[width=0.43\textwidth,height=0.5\textwidth]{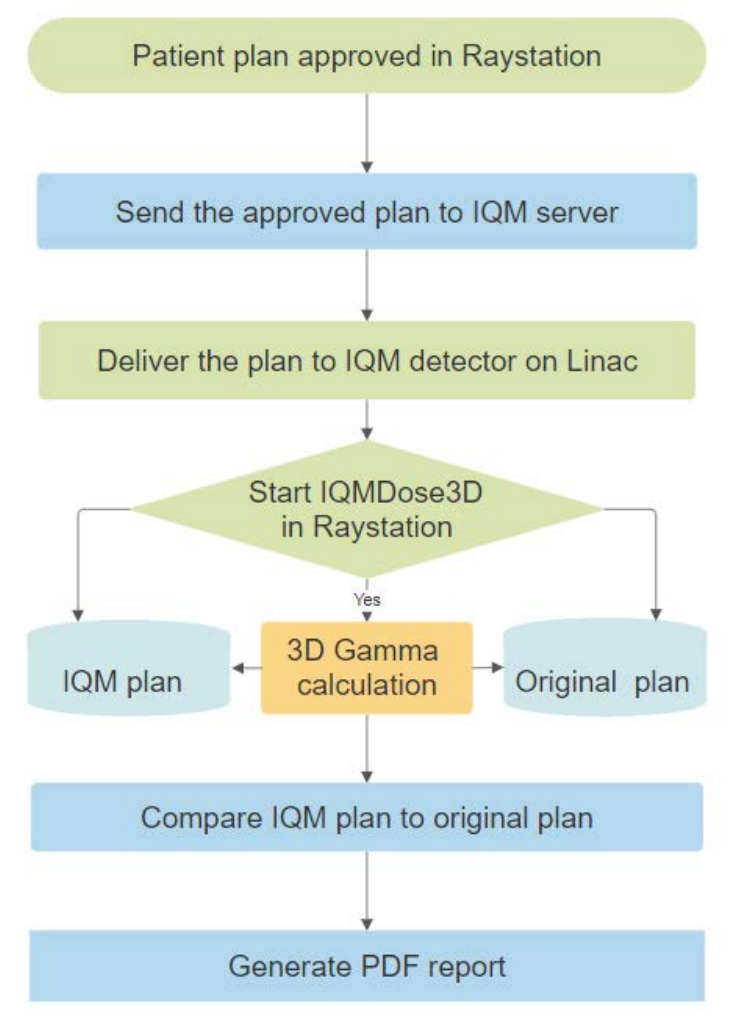}
  \caption{Seamless integration of IQMDose3D into IQM-based PSQA workflow.}
  \label{fig:work_flow}
\end{figure}

\begin{figure*}[h!]
  \centering
  \includegraphics[width=0.6\textwidth,height=0.37\textwidth]{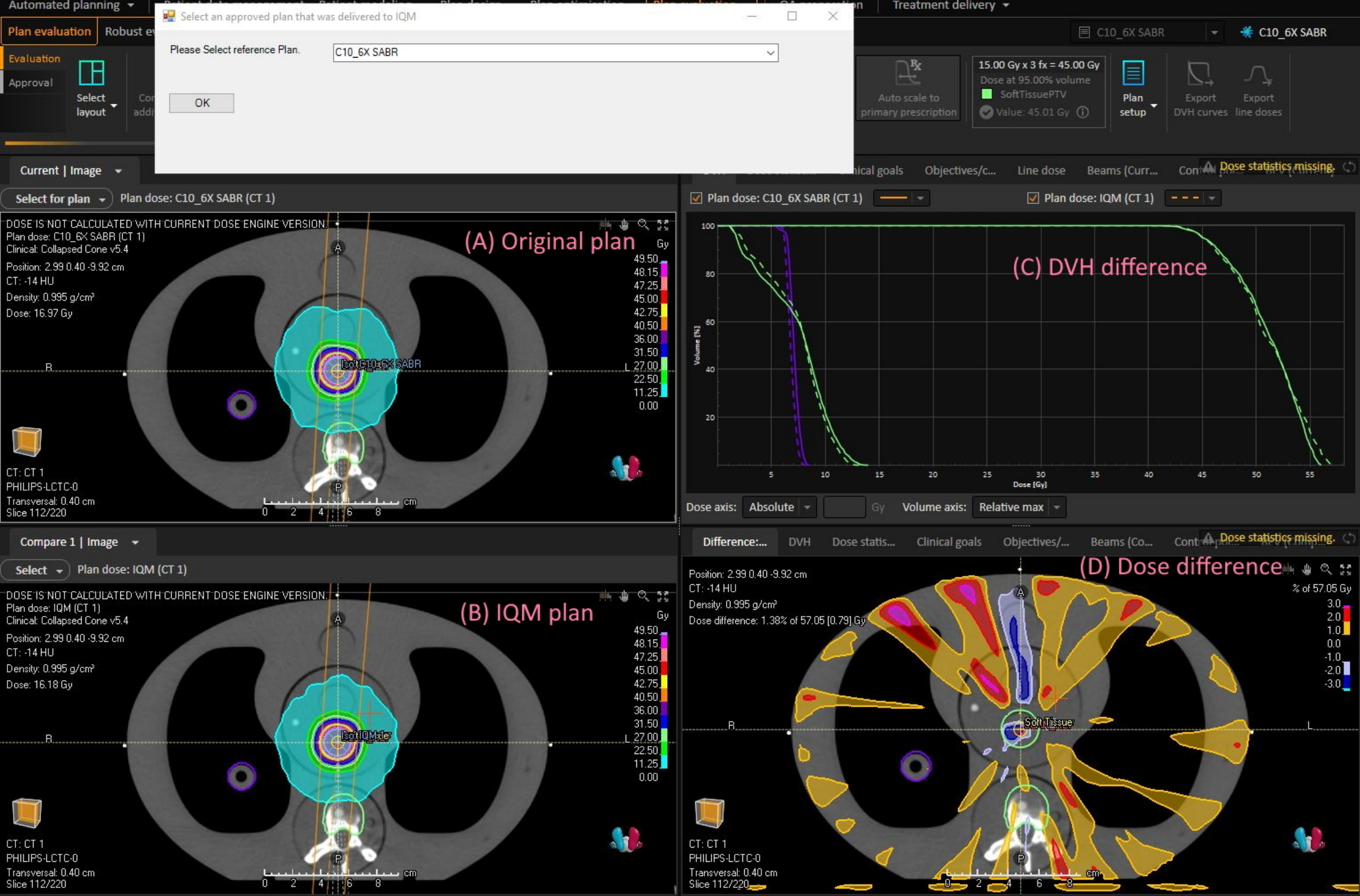}
  \caption{Comparison of the delivered IQM plan to the original plan using 3D dose overlapped with CT images using difference dose images and DVH curves.}
  \label{fig:cmp_dvh}
\end{figure*}

\begin{figure*}[h!]
  \centering
  \includegraphics[width=0.6\textwidth,height=0.36\textwidth]{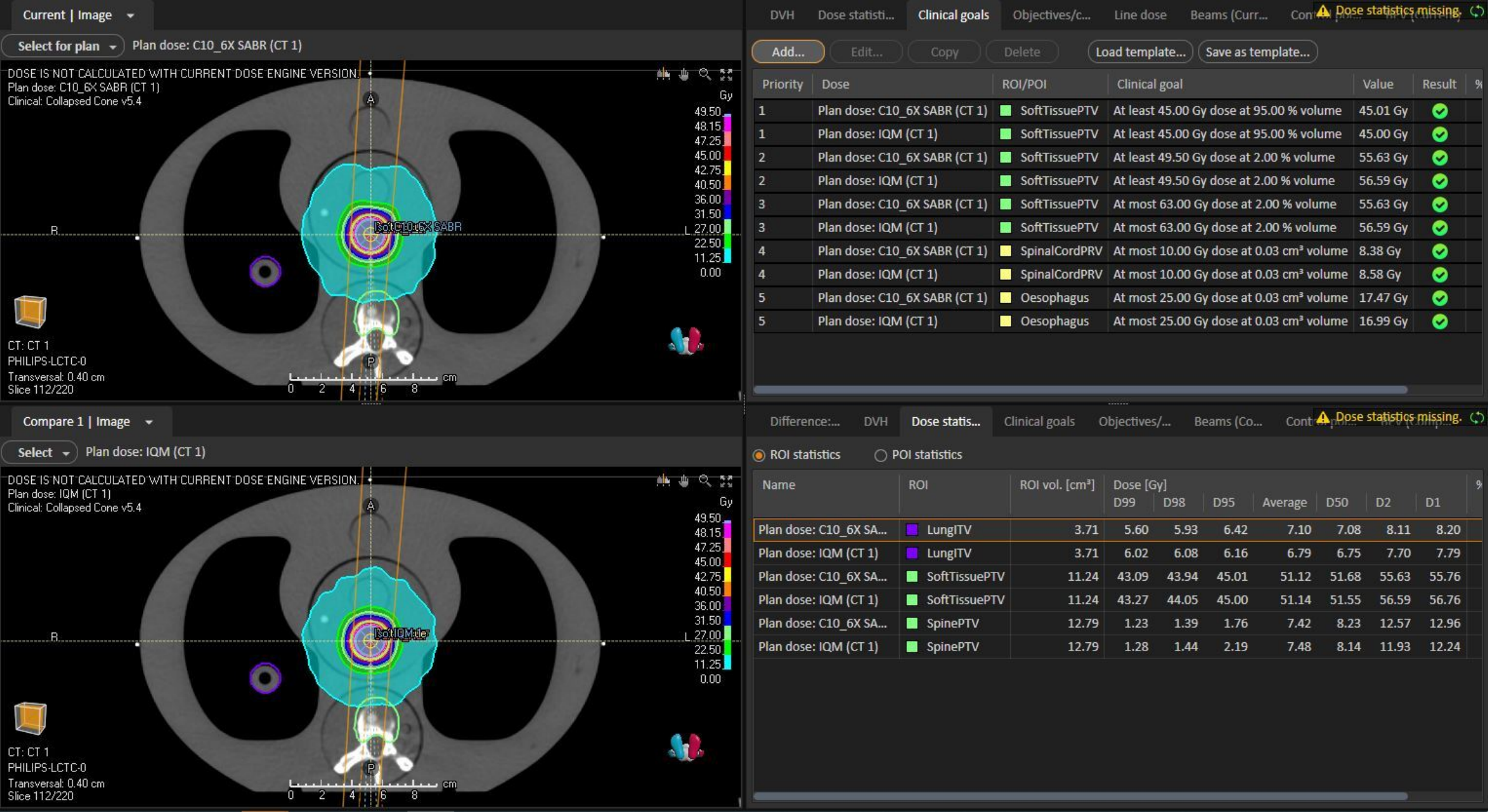}
  \caption{Comparison of the delivered IQM plan (the lower dose image)  to the original plan (the upper dose image) in terms of clinical goals (upper table) and dose statistics (lower table) for PTV, CTV and OARs.}
  \label{fig:cmp_goal}
\end{figure*}

\begin{figure*}[h!]
  \centering
  \includegraphics[width=0.6\textwidth,height=0.37\textwidth]{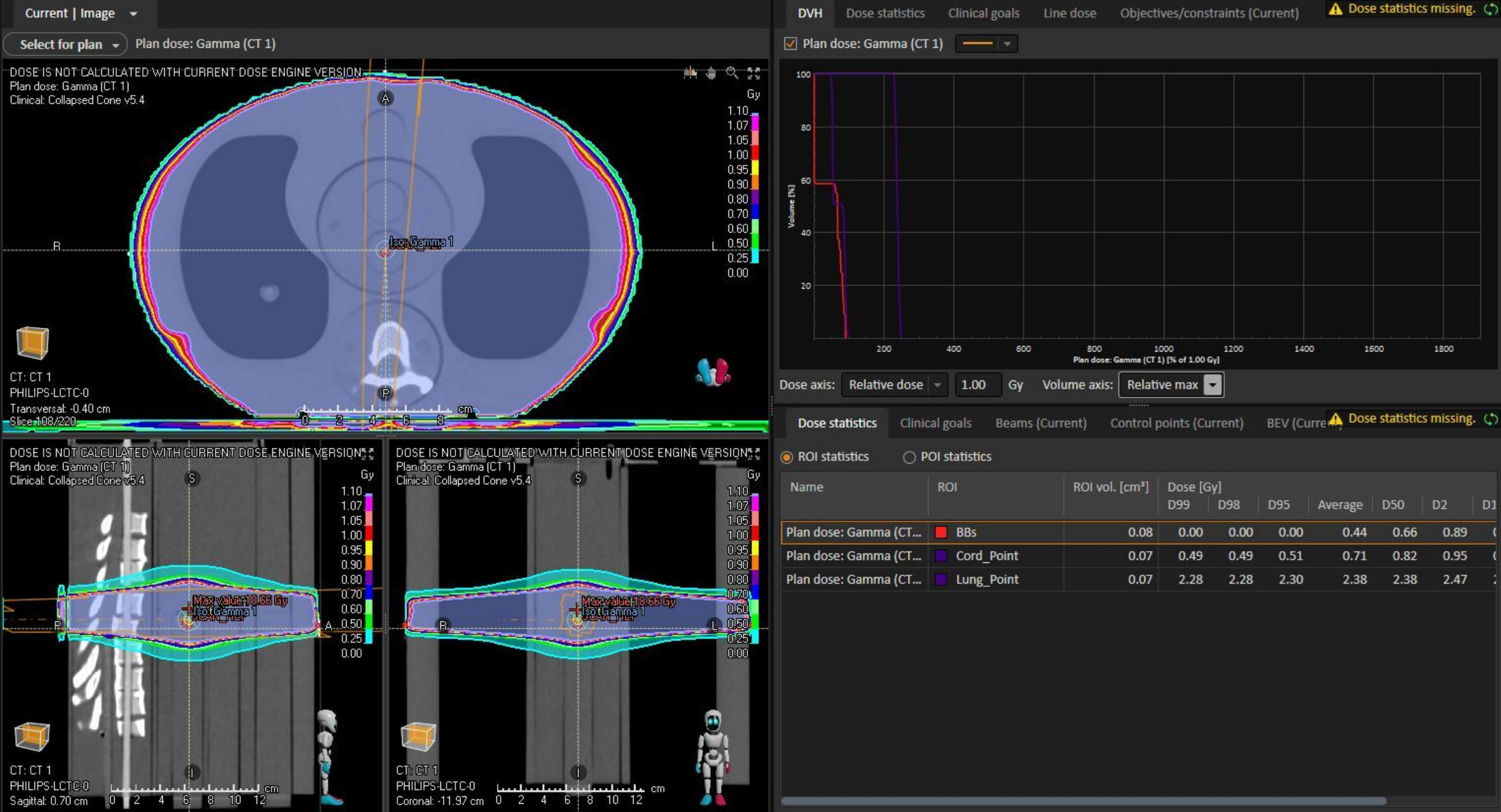}
  \caption{The 3D gamma map overlapped with CT images and structures at transverse, sagittal and coronal planes. The DVH curves are the gamma DVH for different structures. The table shows the gamma statistics for different structures. The gamma map was calculated using the fraction dose with dose criteria of 1\% of the maximum plan dose and distance criteria of 1mm.}
  \label{fig:gamma}
\end{figure*}

\section{Results}

The IQMDose3D-based workflow was shown in Figure \ref{fig:work_flow}. It can be seamlessly integrated into  IQM-based PSQA workflow. Currently, once the clinical plan is delivered to the IQM detector, a PDF report is automatically generated and shows if the patient PSQA is passed or failed based on final accumulated deviation, accumulated and segment-by-segment segment-weighted-pass rate (SPR), which are defined based on the percentage difference between measured and calculated signal \cite{IRTManual}. After the current workflow, the only thing the user needs to do is to open the patient in Raystation and run a script called IQMDos3D, which is the starting point of the tool.

Once IQMDose3D is started, as shown in Figure \ref{fig:cmp_dvh}, a small window will pop up asking the user to select the clinical plan, which was delivered to the IQM detector. After the user clicks OK, the program will automatically generate the delivered IQM plan and Gamma plan containing a 3D gamma map without the user’s interference. The script will pause for the user to review the delivered IQM against the original plan. RayStation provides powerful and flexible tools which can be used to compare the two plans from different aspects.

The PSQA results can be evaluated by directly comparing the dose difference in patient volume as shown in Figure \ref{fig:cmp_dvh}. Three windows showing the doses for original and delivered plans and the difference between them are synchronized. The dose difference can also be viewed slice-by-slice for different patient orientations. The DVH difference for different organs at risk (OARS), planning target volume (PTV) and clinical target volume (CTV) can also be checked by turning them on. 

IQMDose3D-based PSQA workflow also enables users to look at patient QA results from a more clinical perspective. As shown in Figure \ref{fig:cmp_goal}, the user can focus on the clinical goals by looking at changes in the clinical goals of the delivered plan to the original plan. The effect of the delivered IQM plan on the dose statistics on PTV, CTV and OARS can be evaluated by looking at the various indices such as coverage, uniformity and mean or maximum dose for a certain structure as well as dose-volume indices.

As a widely used physics index, a 3D gamma map was also calculated using Raystation’s built-in function. The gamma map was calculated using the fraction dose not the total dose. The Gamma plan containing the 3D gamma map was  overlapped with patient planning CT as shown in Figure \ref{fig:gamma}. Due to the limitation of RayStation scripting, the unit and label of display windows can not be changed, the 3D gamma plan was displayed as dose. The gamma map was calculated using distance criteria of 1mm and dose criteria of 1\% of the max dose of original plans. Except for the overall gamma pass rate, the gamma value can be checked slice-by-slice and structure-by-structure for different patient orientations. The gamma DVH and gamma value statistics for any structure can also be viewed. Instead of using the overall pass rate for the whole patient volume, the gamma pass rate can be calculated for any structure and the range of gamma values within the structure can be visually inspected.

\section{Discussion}

For the current IQM system, the signal was calculated using the patient dicom RT plan file only and what the IQM measured is the charges for the segment and beams. Therefore it essentially just checks the plan transfer between the record and verification (R\&V) system and TPS as well as the deliverability of the patient's plan on the Linac. In contrast, the phantom-based PSQA such as using Electronic Portal Imaging devices (EPID) \cite{Xing2014}  or ArcCheck \cite{Sankar2016} not only verifies the plan transfer and plan deliverability but also verifies the patient plan by comparing calculated dose by TPS and measured dose in phantom. The method and software tool developed in this work filled the gap between current IQM-based PSQA and phantom-based PSQA and made the IQM system become a dose-based system. 

In our centre, currently, we only use IQM for non-high-risk patients. For the patients treated with Stereotactic ablative radiotherapy(SABR) and Stereotactic radiosurgery(SRS), the PSQA was performed using ArcCheck and film. IQMDose3D makes it feasible to use IQM for these patients.  

In comparison to the IQM-based workflow, the significant clinical impact of IQMDose3D-based PSQA is to change the way how the PSQA was evaluated. It lets physicists focus on more the clinical quality indices of the plan in comparison to traditionally gamma-only based PSQA. It can be also evaluated using a gamma map but in a more detailed and flexible way with the help of the Raystation evaluation tool. 

More than 100 centres around the world are currently using IQM for PSQA \cite{IRTManual}. Implementation of the method and proposed PSQA workflow is not limited to RayStation TPS. It can be clinically implemented in these centres despite the TPSs being used. 

On the market, a few commercial products such as  COMPASS from IBA Dosimetry \cite{Sdrolia2015}, and Mobius3D from Varian company \cite{Kim2021} can be used to compare the delivered plan to the original plan using a similar way as described in this work. The delivered plan can be created by these products, but the original clinical plan is created in another TPS. The advantage of the methodology and workflow presented in this work is that the delivered plan and original plan were created in the same system and removed the uncertainties associated with different implementations of two independent systems.

\section{Conclusion}

A software tool was developed for implementing a method proposed to reconstruct the 3D dose in the patient using the signal measured by the IQM detector for patient-specific quality assurance. The software tool can be seamlessly integrated into the IQM-based PSQA workflow.

\printbibliography

@Article{Ghafarian2021,
  author       = {M.Ghafarian, P. Michael, and M.Manuel},
  journal      = {Journal of Applied Clinical Medical Physics },
  title        = {Comparison of pretreatment VMAT quality assurance with the integral quality monitor (IQM) and electronic portal imaging device (EPID)},
  year         = {2021},
  number       = {3},
  pages        = {166},
  volume       = {22},
 
}

@Article{Alharthi2021,
  author       = {T. Alharthi and A. George  and S. Arumugam and  L. Holloway and  D. Thwaites and P Vial. },
  journal      = {Phys Med. },
  title        = {An investigation of the IQM signal variation and error detection sensitivity for patient-specific pre-treatment QA},
  year         = {2021},
  number       = {10},
  pages        = {6},
  volume       = {86},
  doi          = {10.1016/j.ejmp.2021.05.005. Epub 2021 May 25. PMID: 34049118.},
}

@Article{Gray2023,
  author       = {A.  Gray and  A. Xing, V. Moutrie1 V and K. Michel and G. Goozee},
  journal      = {Phys Eng Sci Med},
  title        = {Development of quality assurance procedures for the Integral Quality Monitor (IQM)},
  year         = {2023},
  number       = {},
  pages        = {413},
  volume       = {46},
  doi          = {https://doi.org/10.1007/s13246-023-01228-5. },
}

@Article{Michel2021,
  author       = {K.Michel and A. Gray and A. Xing and G. Goozee and V. Moultrie},
  journal      = {Phys Eng Sci Med},
  title        = {First Australasian experiences of the Integral Quality Monitor (IQM) and the impact of the COVID-19 pandemic on installation},
  year         = {2021},
  number       = {7},
  pages        = {919},
  volume       = {44},
  doi          = {https://doi.org/10.1007/s13246-021-01024-z. },
}

@conference{Xing2023,
    title        = {A Retrospective Study of the Integral Quality Monitor’s Performance for Patient-specific Quality Assurance for Two Commercial Treatment Planning Systems},
    author       = {A. Xing and  A. Gray and  V. Moutrie and K. Michel and  G. Goozee and P.Vial},
    year         = 2023,
    month        = {July},
    booktitle    = {Proceedings of the 65th Annual Meeting of the American Association of Physicists in Medicine (AAPM)},
    publisher    = {Co.Medical Physics Publishing},
    address      = {Denver, USA},
    pages        = {123-127},
    note         = {This is a sample entry for a paper in conference proceedings.},
    editor       = {American Association of Physicists in Medicine (AAPM)},
    organization = {Climate Change Association}
}

@online{IRTManual,
    author = "iRT Systems. Schlossstrasse 1, 56068 Koblenz, Germany. ",
    title = "IQM user reference manual ",
    url  = "www.i-rt.de ",
    addendum = "(accessed: 01.2024)"
}

@online{IRTAlgorithm,
    author = "iRT Systems. Schlossstrasse 1, 56068 Koblenz, Germany. ",
    title = "2018-11-29-User Training-2-Calc Algorithm ",
    url  = "www.i-rt.de ",
    addendum = "(accessed: 01.2024)"
}

@Article{Xing2014,
  author       = {A. Xing and S. Arumugam and S. Deshpande and A. George and L. Holloway and G. Goozee and P. Vial},
  journal      = {In Journal of Physics: Conference Series},
  title        = {Streamlining EPID-based IMRT quality assurance: auto-analysis and auto-report generation},
  year         = {2014},
  number       = {1},
  pages        = {012084},
  volume       = {489},
  doi          = { },
}

@Article{Sankar2016,
  author       = {A. Sankar and A. Xing and T. Yong and  L. Holloway},
  journal      = {Physica Medica },
  title        = {Comparison of three commercial dosimetric systems in detecting clinically significant VMAT delivery errors},
  year         = {2016},
  number       = {10},
  pages        = {1238},
  volume       = {32},
  doi          = { },
}

@Article{Sdrolia2015,
  author       = {A. Sdrolia and KM. Brownsword  and JE. Marsdenand  KT.Alty and CS.Moore  and AW. Beavis},
  journal      = {Physica Medica },
  title        = {Comparison of three commercial dosimetric systems in detecting clinically significant VMAT delivery errors},
  year         = {2015},
  number       = {7},
  pages        = {792},
  volume       = {31},
  doi          = { },
}

@Article{Kim2021,
  author       = {SY. Kim and  J. Park and JW. Park, JW. Yea and SA. Oh},
  journal      = {Progress in Medical Physics  },
  title        = {A Comparison between Portal Dosimetry and Mobius3D Results for Patient-Specific Quality Assurance in Radiotherapy},
  year         = {2021},
  number       = {4},
  pages        = {107},
  volume       = {32},
  doi          = { },
}

\end{document}